\documentclass[a4paper]{article}
\usepackage{INTERSPEECH2021}
\usepackage{multirow}
\usepackage{hyperref}
\newcommand{\ATCO}{ATCO$^2$ }

\title{Detecting English Speech in the Air Traffic Control Voice Communication}
\name{Igor Sz\"oke, Santosh Kesiraju, Ond\v{r}ej Novotn\'{y}, Martin Kocour, Karel Vesel\'{y}, Jan ``Honza'' \v{C}ernock\'{y}}
\address{
  Brno University of Technology, Faculty of Information Technology, Speech@FIT, Czechia}
\email{szoke@fit.vutbr.cz}

\begin{document}

\maketitle
% 

%Passcode 1033X-H4G8H3C3J8

\begin{abstract}
We launched a community platform for collecting the ATC speech world-wide in the \ATCO project. Filtering out unseen non-English speech is one of the main components in the data processing pipeline. The proposed English Language Detection (ELD) system is based on the embeddings from Bayesian subspace multinomial model. It is trained on the word confusion network from an ASR system. It is robust, easy to train, and light weighted. We achieved $0.0439$ equal-error-rate (EER), a $50\%$ relative reduction as compared to the state-of-the-art acoustic ELD system based on x-vectors, in the in-domain scenario. Further, we achieved an EER of $0.1352$, a $33\%$ relative reduction as compared to the acoustic ELD, in the unseen language (out-of-domain) condition. We plan to publish the evaluation dataset from the \ATCO project.

\end{abstract}

\enlargethispage{5mm}

\section{Introduction}\label{sec:Intro}
\ATCO project\footnote{\url{http://atco2.org}} aims at developing a unique platform allowing to collect, organize and pre-process Air-Traffic Control voice communication -- ATC data -- recorded through the VHF channel\footnote{Very High Frequency, Amplitude modulation in range of $108$MHz - $137$MHz}. ATC voice data is tied with a metadata collected through ADS-B channel (altitude, call-sign, GPS coordinates, etc.) and automatically processed. The goal is to create a labeled speech dataset that grows in time, and allows human annotators to correct automatic transcriptions and tag errors. The \ATCO dataset will be useful in research and development of in-cockpit voice enabled applications.

The \ATCO project primarily aims at delivering good quality, low Signal-to-Noise Ratio (SNR) English speech data. Collecting such data is challenging because the data feeders\footnote{Volunteers with antennas and SDR devices capturing the communication from the air and uploading to \ATCO servers.} are spread globally, and the data contains many local languages and dialects of English. As we do not have control over the recording conditions, the quality of the data may vary\footnote{As we are launching the platform now (March 2021), we present our findings on data collected from Prague (LKPR) and Brno (LKTB) airports in Czechia -- our bootstrap data feeders. On the other hand, we do not expect significant change in the data properties and quality worldwide.}

The ATC data is challenging for automatic speech processing~\cite{lin_sichuan_atcnlu_2021}, especially in our environment driven by volunteers:
\begin{itemize}
  \item Various noise conditions. The majority of data is ``clean'' ($82$\% of data has SNR above $10$dB), but there are dialogues where the SNR difference between conversation sites is very large.  \vspace{-1mm}
  \item Strongly accented English. The speakers’ English accent varies widely. From native speakers (pilots), through international accents (French, German, Russian, etc.) (pilots and ATCOs) to the strong local accent (Czech pilots and ATCOs in our initial data).\vspace{-1mm}
  \item Mixed words and phrases. The Czech ATCOs vocabulary is a mix of Czech and English words. They use standard greetings in Czech which can be a significant portion of an English sentence if the command is short. On the other hand, they use many English words (alphabet, some commands) in Czech sentences. Moreover they use a significant amount of ``Czenglish'' words. Our international partners report similar situation in their countries.
\end{itemize}

Because of the sparsity in VHF data (speech was present in $1\% - 8\%$ of time), and local language environment (English language was present in $70-80\%$ of the segments), one VHF channel can generate about $20-30$ hours of English speech per month.

\subsection{Motivation}

\begin{figure}[!t]
  \centering
  \includegraphics[width=\linewidth]{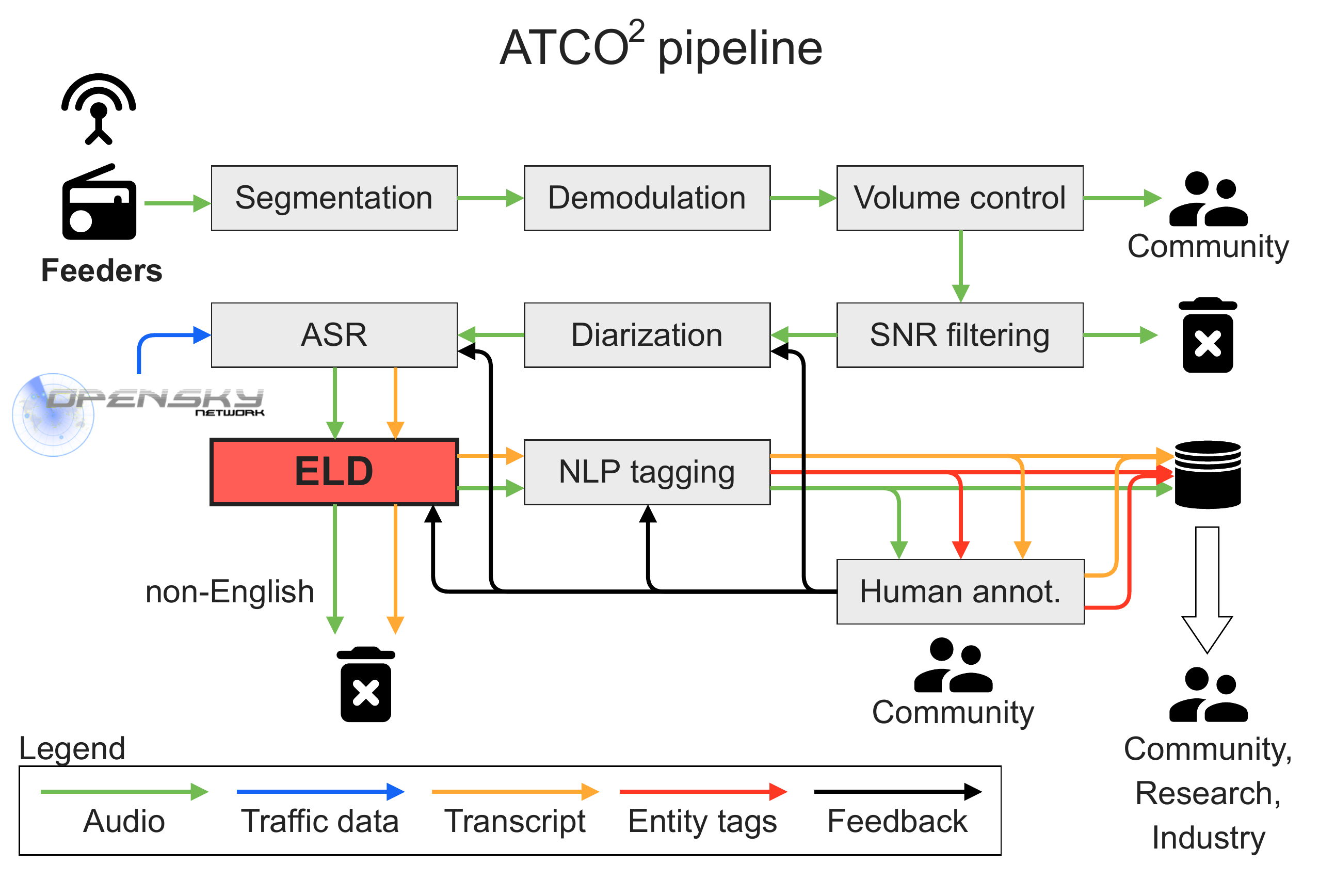}
  \vspace{-7mm}
  \caption{\ATCO data flow pipeline. The ELD block is used to filter our non-English speech. The automatically processed data are then annotated and checked by humans. Their feedback is used to adapt particular blocks using active learning approaches.}
  \label{fig:atco_pipeline}
  \vspace{-3mm}
\end{figure}

The \ATCO data processing pipeline should continuously process large amounts of data in an on-line manner (see Figure~\ref{fig:atco_pipeline}). The decision on (non)English speech should be light-weight and robust. It has to cope with the above mentioned channel, acoustic and lexical challenges while preserving an easy way to update a model to newer airports and countries.

We decided to explore the English language detection (ELD) based on an ASR output for the following two reasons: 1) there is about $20$\% of non-English speech, so running ASR on all the data does not bring a large overhead (compared to filtering non-English prior to the ASR block), 2) standard acoustic-based Language IDentification (LID) can be too data-consuming~\cite{FITPUB9754, 8462403, LIDXvec} and its accuracy suffers in bi-lingual code-switching environments~\cite{chandak2020streaming}. The ASR-based ELD systems share similarities with phonotactic LID sytems~\cite{Zissman:1995:Ph_LID,Lee:2020:sub_LR,Mehdi:2011:SMM, Mehdi:2013:SnGM}: the former relies on word confusion networks, the latter relies on phoneme sequences.

The most recent works targeting similar problems as ours are related to the AI-powered smart voice assistants. Here the motivation is to detect the language of short commands efficiently. The challenge increases in bilingual or multilingual environments with code switching. 
\cite{wang2019signal,chandak2020streaming} improved language detection in multilingual speech recognition for voice assistants: \cite{wang2019signal} experimented with combination of acoustic LID with ASR features such as $1$-best, confidence, AM and LM scores. They were able to improve the language detection by $52\%$ when adding ASR-based features to the acoustic LID. The motivation of \cite{chandak2020streaming} was to improve language detection in bilingual environment. They used combined acoustic and text (ASR $1$-best) based features to detect language of a spoken command. They achieved about $56\%$ relative error rate improvement using text-based LID compared to acoustic one and another $5\%$ improvement with a combination of acoustic and text based features post-processed by a DNN.

The ATC domain is similar to voice assistants in the sense of noisy channel, bilingual environment (English and the local language) and code switching. On the other hand, we do not require a real time processing in our case.

This paper is aimed at the English Language Detection (ELD) in ATC domain. We describe an effective way of filtering out non-English language. We evaluate on our dataset described in Section~\ref{sec:Data}. The description of acoustic ELD based on state-of-the-art x-vectors is given in Section~\ref{sec:Xvec}. As the speech-to-text technology is one of our building blocks, we briefly discuss it in Section~\ref{sec:ASR}, we kindly ask the reader to follow~\cite{KocourIS2021} for more information. Description of various language detection systems based on the ASR output is presented in Section~\ref{sec:NLP}. The experiments and results are outlined in Section~\ref{sec:exp}, and we conclude with directions for future work in Section~\ref{sec:conclusion}.

\enlargethispage{5mm}

\section{Datasets}\label{sec:Data}

We found only one recent bi-lingual ATC dataset -- English / Chinese ATCSpeech~\cite{Yang2020ATCSpeechAM}\footnote{We have not found any suitable way how to acquire the corpus at the time of writing this paper.}. Other datasets are English language only~\cite{AIRBUS, LDC_ATCC, ATCOSIM, PILSEN_ATC_LREC2018,PILSEN_ATC_LRE2019,srinivasamurthy_semisupervised_2017} or multilingual but not in ATC domain~\cite{N4NATO}. 

Regarding the lack of multilingual ATC datasets with language tagging, we had to create one ourselves based on the collected \ATCO data. At the time of writing this paper, we had coverage of Czechia (Prague and Brno airports. i.e. English / Czech). To test robustness of our solution, we used HAAWAII project data from Iceland (Keflav\'{i}k airport, i.e. English / Icelandic) as unseen (out-of-domain) data.

Assigning the language tag to particular segment (a short dialog between ATCO and pilot)  was mainly based on the language of spoken numerals. We used the following rules:
\begin{itemize}
    \item The default tag is English.\vspace{-1mm}
    \item Language of greetings is ignored.\vspace{-1mm}
    \item If numerals are spoken in Czech / Icelandic, the tag is Czech / Icelandic respectively.\vspace{-1mm}
    \item If there is ``out of ATC grammar'' speech in Czech / Icelandic, the tag is Czech / Icelandic.\vspace{-1mm}
    \item If no numerals are present and the commands / values are clearly in English, the tag is English.\vspace{-1mm}
    \item Otherwise, discard the speech segment and tag it as mixed (clearly both languages are present) or unknown language (cannot be determined).
\end{itemize}

The \textbf{\ATCO ELD dataset} is formed from two airports and comprises two subsets (see Table~\ref{tab:ATCOdataset}). Both subsets were recorded during a period of several months, and none of the recordings overlap in time.

\begin{table}[th]
  \caption{\ATCO ELD datasets. The first line is the Number of segments / Total length of speech [minutes]. LKPR -- Czechia, Prague airport, LKTB -- Czechia, Brno airport}
% The second line is the Average SNR [dB]\cite{Chanwoo:SNR:IS2008}.  
  \label{tab:ATCOdataset}
  \centering
  \vspace{-2mm}
\begin{tabular}{|c|c|c|c|c|} 
\hline
\multirow{2}{*}{} & \multicolumn{2}{c|}{LKPR} & \multicolumn{2}{c|}{LKTB}  \\ 
\cline{2-5}
     & EN     & CZ      & EN     & CZ       \\ 
\hline
ENCZ-Dev &  596/77 & 492/61 & 579/66 & 453/61 \\
\hline
ENCZ-Eval & 820/128 &   308/55 &  515/57 & 289/36  \\
\hline
\end{tabular}
\vspace{-2mm}
\end{table}

The \textbf{HAAWAII ELD dataset} is formed from a single airport and consists of two subsets (see Table~\ref{tab:HAAWAIIdataset}). Both subsets were recorded during a period of two months and none of the recordings overlap in time.

\begin{table}[th]
  \caption{HAAWAII ELD datasets. The first line is the Number of segments / Total length of speech [minutes]. BIKF -- Iceland, Keflav\'{i}k Airport}
  \label{tab:HAAWAIIdataset}
  \centering
  \vspace{-2mm}
\begin{tabular}{|c|c|c|} 
\hline
\multirow{2}{*}{} & \multicolumn{2}{c|}{BIKF} \\ 
\cline{2-3}
     & EN     & IS      \\ 
\hline
ENIS-Dev & 404/69 & 109/19        \\
\hline
ENIS-Eval & 414/80 & 111/24 \\
\hline
\end{tabular}
\vspace{-2mm}
\end{table}

We plan to enlarge the \ATCO database and add more languages during the project. We intend to publish the datasets during $2021$. The HAAWAII dataset is proprietary and the project agreement does not allow to publish it.

\section{Baseline Acoustic Based English Language Detection}\label{sec:Xvec}

Our baseline acoustic ELD system is based on neural network x-vector technology, initially introduced for speaker verification~\cite{Snyder2017DeepNN}. NN as an x-vector extractor was shown to be very robust in noisy environments~\cite{snyder2018x-vectors, novotny2018x-vectors}. The original x-vector topology was successfully applied as a language recognizer~\cite{LIDXvec}. 

Our system replaces the Time Delay Neural Network  (TDNN) topology with a more efficient topology based on ResNet~\cite{ResNetOrigin} with $18$ layers and  $64$ filter bank channels as the acoustic features. Softmax was used as the output layer non-linearity with a class for each language. 

Our training set contains data from NIST LRE 2009--2017~\cite{NISTLRE2009plan,NISTLRE2011plan,NISTLRE2015plan,NISTLRE2017plan}, Babel~\cite{babel}, NIST SRE16~\cite{NISTSRE2016plan}, and Fisher English~\cite{fisher}. The data was augmented using Kaldi~\cite{Kaldi} pipeline~\cite{FITPUB12039}, without the babble noise branch. The training data comprises of $62$ languages with $644$k segments, totalling $21.6$k hours in duration.

The x-vector embeddings are extracted and fed into Gaussian linear classifier (GLC) for English and non-English classification. Experimental details are given in Section~\ref{sec:exp}.

\section{Speech-to-text}\label{sec:ASR}
The English ASR was trained on $7$ corpora: AIRBUS~\cite{AIRBUS}, HIWIRE~\cite{HIWIRE}, LDC ATCC~\cite{LDC_ATCC}, MALORCA~\cite{srinivasamurthy_semisupervised_2017}, N4 NATO~\cite{N4NATO}, ATCOSIM~\cite{ATCOSIM}, UWB ATCC~\cite{PILSEN_ATC_LREC2018,PILSEN_ATC_LRE2019}. The total duration of audio was $136.3$ hours.

We used hybrid speech-to-text recognizer based on Kaldi~\cite{Kaldi}, trained with Lattice-free MMI training~\cite{povey_lfmmi_IS2016}. The neural network had $6$ convolutional (\texttt{conv-relu-batchnorm-layer}) layers, followed by $9$  semi-orthogonal components (\texttt{tdnnf-layer})~\cite{TDNN-F}, and the model has two pre-final and output layers. In total, the model has $12.93$ million trainable parameters, and the number of left bi-phone tied-states is $1680$. The input features are high-resolution Mel-Frequency Cepstral Coefficients  (MFCC) with online Cepstral Mean Normalization (CMN). The features are extended with online i-vectors~\cite{peddinti_aspire_ivector_ASRU2015,saon_ivector_ASRU2013}.

We used a vocabulary of $28.4$k unique tokens. Out of this, $15.3$k tokens are $5$-letter way-points from~\cite{olive_traffic_OSS2019}, and $5.2$k tokens are airline designators for call-signs\footnote{\url{https://en.wikipedia.org/wiki/List_of_airline_codes}}$^{,}$\footnote{\url{https://www.faa.gov/documentLibrary/media/Order/7340.2J_Chg_1_dtd_10_10_19.pdf}}. Pronunciations were generated using Phonetisaurus tool~\cite{phonetisaurus} with a G2P model trained on Librispeech lexicon~\cite{panayotov_librispeech_ICASSP2015}. Standard $3$-gram Language Model (LM) was obtained by interpolating LMs from individual training corpora. 
The ATC speech should follow the standard~\cite{icao_radiotelegraphy_annex10_vol2_2001}, 
but ``free speech'' can also appear in the data.

The English ASR achieved $8.4\%$ Word Error Rate (WER) on our custom Airbus-dev dataset~\cite{AIRBUS} and $8.9\%$ WER on MALORCA project -- Vienna airport test set~\cite{srinivasamurthy_semisupervised_2017}. The ASR achieved $20.4$\% WER on $5$ minutes English subset of our ENCZ-Eval data\footnote{We are on the beginning of the transcription process so we have only $5$ minutes of transcribed data so far. The provided WER cannot be taken as very reliable. On the other hand, it provides an idea about how well the ASR works.}

\enlargethispage{5mm}

To the best of our knowledge, there exists no Czech ATC speech corpus for training an in-domain ASR. Therefore, we relied on an off-the-shelf Czech ASR, which is $8$kHz ASR system trained on a mixture of $2512$ hours of telephone and distant microphone speech, including data augmentation. We used a hybrid speech recognizer trained with Kaldi. We used TDNN architecture~\cite{peddinti:IS:2015:tdnn}, containing $7$ layers each with $450$ neurons. The NN was trained with Lattice-free MMI objective, sub-sampling by factor of $3$ and bi-phone targets as suggested in~\cite{povey_lfmmi_IS2016}. The features were $40$-dimensional Mel-filter bank outputs. We added only the local way-points, the airline designators for call-signs, and a hand-filtered set of alphabet and commands~\cite{icao_radiotelegraphy_annex10_vol2_2001} to the vocabulary. This word list was mixed with the original LM with an ad-hoc $0.3$ weight. The ASR achieved $19.1$\% WER on prompted Czech speech recorded on distant microphone in noisy environment and $30.7$\% WER on telephone speech conversations.

The ASR achieved $79.2$\% WER on $2$ minutes of Czech subset from our ENCZ-Eval data. The high WER shows that the ASR is not adapted to the ATC domain. Since we are in the beginning of the transcription process, we have only $2$ minutes of transcribed data so far. But, we can see that Czech ASR system has much higher WER ($79.2$\%) compared to the English ASR system ($20.4$\%) on our ATC data.

A more detailed description of the ASR systems is out of the scope of this paper. The reader is kindly asked to find the details in~\cite{KocourIS2021}. 

\section{ASR-based English Language Detection}\label{sec:NLP}

As described in the Sections~\ref{sec:Intro} and~\ref{sec:Data}, the spoken utterances are code-mixed, i.e., each utterance may contain a greeting in one language (e.g. Czech) followed by a message in a different language (e.g. English). We hypothesize that the spoken words in the message context could be helpful in determining the \textit{significant} language of the utterance. Hence we employed models that are prominent in text modelling / classification.

For each utterance (or recording), we extract bag-of-words statistics from the transcriptions automatically generated by the ASR. More specifically, we use both English and Czech ASR systems to generate confusion networks for every recording 
(confusion network is a sausage-like structure with bins containing soft counts or probabilities for a list of candidate words in each bin).  The word statistics for each utterance are accumulated to form an utterance-by-word count matrix. The vocabulary sizes of EN and CZ are $27675$ and $326509$ respectively. We filtered out words whose total occurrence in training set is less than $1e-03$, which resulted in $149673$ unique words. We can observe that the vocabulary size is much higher than the number of utterances (Table~\ref{tab:ATCOdataset}).

The filtered utterance-by-word counts are used by the Bayesian subspace multinomial model (BaySMM)~\cite{Kesiraju:2020:BaySMM} to learn utterance embeddings\footnote{\texttt{\url{https://github.com/BUTSpeechFIT/BaySMM}}}. The model is trained in an unsupervised way, and can learn to represent utterance embeddings ($\mathbf{u}$) in the form of Gaussian distributions, thereby encoding the uncertainty in the covariance matrix, i.e.,
     $p(\mathbf{u}) = \mathcal{N}(\mathbf{u} \mid \boldsymbol{\nu}, \mathrm{diag}(\boldsymbol{\gamma})^{-1})$, where $\boldsymbol{\nu}$ represents the mean of the utterance embedding and $\boldsymbol{\gamma}$ represents the diagonal precision (or the uncertainty).
The estimated uncertainties  are high for shorter utterances with more unique words as compared to longer utterances with many tokens. Modelling the uncertainties helps in training a robust classifier for downstream tasks.

\begin{table}[t]
    \caption{Hyper-parameters of Bayesian SMM. $\mathrm{so}$ stands for semi-orthogonal constraint.}
    \label{tab:hyper}
    \vspace{-2mm}
    \begin{tabular}{ll} \toprule 
     {Hyper-parameter} & {set / range} \\ \midrule
     regularization type & $\{\ell_1, \ell_2, \mathrm{so}\}$ \\
     regularization weight & $\{1e-05, 2e-05, \ldots, 1e-02\}$ \\
     subspace dimension & $\{32, 64, \ldots, 160\}$ \\ \bottomrule
    \end{tabular}
    \vspace{-4mm}
\end{table}

The primary classifier for ELD is the generative Gaussian linear classifier with uncertainty (GLCU)~\cite{Kesiraju:2020:BaySMM,Sandro:2015:IS}, which is trained on the utterance embeddings $p(\mathbf{u})$ obtained from BaySMM. Additionally, we also train a binary logistic regression for comparison.

Our baseline ASR-based ELD system relies on \textit{term frequency inverse document frequency} (TF-IDF) representation~\cite{Manning:SNLP} of utterance statistics followed by a binary logistic regression for English / non-English classification.

\section{Experiments and results}
\label{sec:exp}
Since our aim is to filter-out non-English and retain as much English data as possible, we chose equal-error-rate (EER) as our evaluation metric.

We used ENCZ-Dev from Table~\ref{tab:ATCOdataset} and ENIS-Dev from Table~\ref{tab:HAAWAIIdataset} as the training sets for unsupervised learning in BaySMM. Unless and otherwise mentioned, the utterance-by-word matrix is obtained by pooling word statistics from the confusion networks of both EN and CZ ASR systems. No additional, external data was used in training the BaySMM. The embeddings are extracted for all  utterances from all the sets. We performed $5$-fold cross-validation classification experiments on the training set to choose the best model, i.e., hyper-parameter configuration (Table~\ref{tab:hyper}). The training details of x-vector based system is described in Section~\ref{sec:Xvec}.

Table~\ref{tab:eer_asr_xvec} presents the EER results for the baseline acoustic ELD based on x-vectors, and the proposed ASR-based ELD sytems. The training data for classifier is the union of ENCZ-Dev and ENIS-Dev sets and the evaluation is done independently on ENCZ-Eval and ENIS-Eval sets. From Table~\ref{tab:eer_asr_xvec}, we see that all the ASR-based approaches outperform acoustic x-vector system. Further, the BaySMM has an EER of $0.0439$, a $50\%$ relative reduction as compared to the x-vector system. The major gains in ASR-based system are due to fact that the channel variability/noise in the data is addressed by the ASR and hence text embeddings and the classifier are not affected. On the other hand, the acoustic ELD has a more difficult task of dealing with the channel variability / noise, while aiming to classifying the acoustic features to the target language.

\begin{table}[t]
\centering
\caption{EER results on Eval sets using acoustic and ASR-based approaches. Training data are combined ENCZ-Dev and ENIS-Dev sets. $\mathrm{dim}$ refers to dimension of feature representation for the classifier (CLF).}
\label{tab:eer_asr_xvec}
\vspace{-2mm}
\scalebox{0.95}{
\begin{tabular}{lrlrr} \toprule
 Model    & $\mathrm{dim}$ & CLF & ENCZ-Eval & ENIS-Eval \\ \midrule
 x-vector & 256          & GLC        &  0.0870  &  0.1450  \\
 BaySMM   & 64           & GLCU       &  \textbf{0.0439} & \textbf{0.1003} \\
 BaySMM   & 160          & LR         &  0.0467 & \textbf{0.1003} \\
 TFIDF    & 149673       & LR         &  0.0538 & 0.1084 \\
 \bottomrule
\end{tabular}
}
\vspace{-2mm}
\end{table}

\subsection{Effect of training data from target language / airport}
Table~\ref{tab:eer_train_set} presents the results for an out-of-domain scenario, i.e., the classifier is trained only on ENCZ-Dev,  and we evaluate on ENCZ-Eval and ENIS-Eval sets. This gives us an idea of how well systems perform on data from an unseen language / airport. We can see from the Table~\ref{tab:eer_train_set} that ASR-based approaches achieve lower EERs as compared to the acoustic system. The secondary language / airport from ENIS-Eval set is totally unseen in the training data, and we can see that ASR-based system has reduced the EER by $35\%$ relative compared to the acoustic ELD.
Having training data from target airport (ENIS-Dev) reduces the EER from $0.1352$ to $0.1003$, a $25\%$ relative reduction.

\begin{table}[t]
    \centering
    \caption{EER results on Eval sets using acoustic and ASR-based approaches. Training data is ENCZ-Dev only.} 
    \label{tab:eer_train_set}
    \vspace{-2mm}
    \scalebox{0.95}{
    \begin{tabular}{lrlrr} \toprule
    Model & $\mathrm{dim}$ & CLF & ENCZ-Eval & ENIS-Eval \\ \midrule
    x-vector & 256    & GLC  & 0.0890          &  0.2030  \\
    BaySMM   & 64     & GLCU & 0.0435          & \textbf{0.1352} \\
    BaySMM   & 128    & LR   & \textbf{0.0423} & 0.1807 \\
    TFIDF    & 108000 & LR   & 0.0554 & 0.1888 \\
    \bottomrule
    \end{tabular}
    }
    \vspace{-4mm}
\end{table}

\subsection{English vs target-language ASR systems}
We also analyse the effect of individual ASR systems for the ASR-based ELD.
Table~\ref{tab:eer_en_cz_asr} shows the EER results for English and Czech ASR-based ELD systems. We can see (column ENCZ-Eval) that having only a matching secondary language (CZ) ASR helps to reduce the EER ($0.0905 \rightarrow 0.0621$) as compared to having only an English ASR. However, having only a mismatched (CZ) ASR (column ENIS-Eval) increases the EER ($0.1096 \rightarrow 0.1807$), as compared to having only an English ASR. We plan to conduct  a more detailed analysis as we collect more data from multiple airports / languages.

\begin{table}[t]
    \centering
    \caption{EER results on Eval sets using individual ASR-based approaches. Training data are combined ENCZ-Dev and ENIS-Dev sets.} 
    
    \label{tab:eer_en_cz_asr}
    \vspace{-2mm}
    \scalebox{0.86}{
    
    \begin{tabular}{llrlrr} \toprule
    ASR & Model  & $\mathrm{dim}$ & CLF  & ENCZ-Eval & ENIS-Eval \\ \midrule
    \multirow{3}{*}{EN} & BaySMM & 64     & GLCU      & 0.0937 & 0.1352 \\
                        & BaySMM & 160    & LR        & \textbf{0.0905} & \textbf{0.1096} \\
                        & TFIDF  & 27675  & LR        & 0.1008 & 0.1165 \\ \midrule
    \multirow{3}{*}{CZ} & BaySMM & 64     & GLCU      & \textbf{0.0621} & 0.2331 \\
                        & BaySMM & 128    & LR        & 0.0621 & \textbf{0.1807} \\
                        & TFIDF  & 130262 & LR        & 0.0850 & 0.2331   \\
    \bottomrule
    \end{tabular}
    }
    \vspace{-3mm}
\end{table}

\section{Conclusions}
\label{sec:conclusion}
We presented our continuing efforts in collecting ATC speech data, which holds future for in-cockpit voice enabled applications. We  described the need for filtering non-English speech, an important pre-processing step for downstream applications relying on the ATC speech data. We explored both acoustic and ASR-based approaches for detecting English utterances from the collected ATC data. Our results show that ASR-based English Language Detection (ELD) systems are robust and achieve low equal-error-rates as compared to acoustic ELD systems based on x-vectors. We have also shown that the ASR-based ELD can be applied to the new airport with a different secondary language, and achieves $0.1352$ EER. This can be of great help as one can filter out non-English with minimal resources and an off-the-shelf secondary ASR. Moreover, the ASR-based ELD only requires to have a robust English ASR (secondary ASR gives additional improvements), whereas the acoustic ELD needs to be trained on multiple languages under consideration. Furthermore, ASR is a major component in downstream applications on ATC speech data, hence using the same ASR output for  filtering out non-English is more suitable and efficient. We note that both acoustic ELD and ASR systems are data hungry on their own, whereas the embeddings from ASR and the classifier are not. Hence, retraining or adapting a text embedding model such as Bayesian subspace multinomial model will be relatively fast.

In future, we plan to collect more data from more languages and airports. More data will also enable experimenting with fusion of both approaches. As part of the \ATCO project, we plan to make the data public. We also plan to test our approach on the ATCSpeech Chinese / English speech data if available.

\section{Acknowledgements}
The work was supported by European Union’s Horizon 2020 projects No. 864702 - \ATCO and No. 884287 HAAWAII. 

\bibliographystyle{IEEEtran}

\bibliography{ATCO_LID}

\end{document}